\documentstyle[prl,twocolumn,aps]{revtex}
\input psfig
\begin{document}
\draft

 \newcommand{\mytitle}[1]{
 \twocolumn[\hsize\textwidth\columnwidth\hsize
 \csname@twocolumnfalse\endcsname #1 \vspace{1mm}]}

\mytitle{
\title{Renormalization group study of the 1d polaron problem}

\author{Markus Keil$^1$ and Herbert Schoeller $^{2,3}$}

\address{
$^1$ Institut f\"ur Theoretische Physik,
Universit\"at G\"ottingen, 37073 G\"ottingen, Germany\\
$^2$Forschungszentrum Karlsruhe, Institut f\"ur 
Nanotechnologie, 76021 Karlsruhe, Germany\\
$^3$Institut f\"ur Theoretische Festk\"orperphysik, Universit\"at
Karlsruhe, 76128 Karlsruhe, Germany}

\date{\today}

\maketitle

\begin{abstract}
We investigate the one-dimensional polaron problem and calculate
the ground state energy and the effective mass. We use a real-time
renormalization group method and compare our results with
first and second order perturbation theory, with
Feynman's variational principle and with the method 
of Lee, Low and Pines.
\end{abstract}
}

{\it Introduction}.
The polaron has often been examined since Fr\"ohlich proposed the
corresponding Hamiltonian \cite{froehlich1,review1}. It serves as a standard
model for various problems involving a nonrelativistic particle moving
in a scalar field, e.g. the interaction between nucleons and scalar mesons
or a single electron in a solid interacting with longitudinal optical
phonons. The physical picture is that the particle polarizes the
environment and must drag this polarization with it, which affects
its energy and effective mass. To describe the polaron not only in the
weak coupling limit one has to consider it beyond perturbation theory.
The most successful methods for determining the ground state energy
and effective mass of the polaron are those of Lee, Low and Pines
\cite{leelowpines,leepines}, Pekar et al. \cite{pekar1,pekar2}, and 
Feynman \cite{feynman}. Since polaron effects have been 
observed in low-dimensional systems, the problem has also been studied 
in two dimensions \cite{sak,sarma,devreese}.

In this paper we will examine the one-dimensional polaron problem.
It can be realized e.g. for a Bloch-electron in a one-dimensional wire 
or macromolecular structure. The excitation of an electron will be 
strongly influenced by the interaction with optical phonons 
\cite{review2}. If the conduction band is partially filled, one
can linearize the electronic spectrum and the model is exactly
solvable by using bosonization techniques \cite{meden-etal}.
However, if the conduction band is empty, it is 
necessary to consider a quadratic spectrum for the electron
with a bare mass $m_0$. This leads to the one-dimensional Fr\"ohlich
Hamiltonian with a constant coupling to the phonons.
It is our purpose to examine the ground state energy and the
effective mass of the electron for this problem. Thereby we will
use a recently developed real-time renormalization group (RG)
technique \cite{schoeller1}. While this method actually allows 
nonequilibrium descriptions \cite{schoeller2}, in the present paper 
we have only used it to determine spectral properties.
Furthermore, we will compare our results with first
and second order perturbation theory, with Feynman's variational
principle, and with the method of Lee, Low and Pines generalized to
the one-dimensional case with finite band width. For not too large 
coupling constants we find a ground state 
energy near Feynman's method and a value for the effective mass 
between the result of Feynman and the one of Lee, Low and Pines.

{\it Model}.
The polaron problem is modelled by the Fr\"ohlich Hamiltonian
$H=H_0+H_1$ with 
\begin{eqnarray}
H_0 &=& \sum_k\epsilon_k\, c^{\dagger}_kc_k + \sum_q\hbar\omega_q\,
a^{\dagger}_qa_q \label{H0eq}\\ 
H_1 &=& \sum_{k,q}M_k^q \, (a^{\dagger}_{-q} + a_q) c^{\dagger}_{k+q}c_k\,.
\label{H1eq}
\end{eqnarray}
$c^{\dagger}_k(c_k)$ is the creation (annihilation) operator of the
electron, its energy $\epsilon_k$ is given by the electron dispersion
$(\hbar k)^2/2m_0$, $m_0$ being the bare mass of the electron in the
conduction band and $a^{\dagger}_q(a_q)$ creates (annihilates) a phonon with
frequency $\omega_q$. Since the interaction is dominated by the
longitudinal optical branch, we presume dispersionless phonons,
i.e. $\omega_q=\omega$. While the electron-phonon interaction
coefficients $M^q_k\equiv M^q$ are proportional to $1/q$ in case of the bulk
polaron, the one-dimensional situation involves a $q$-independent
coefficient $M$ \cite{devreese}. We define
\begin{displaymath}
M =
(\frac{4m_0\omega}{\hbar})^{\frac{1}{4}}\frac{\sqrt{\alpha}}{\sqrt{L}}\,,
\end{displaymath}
where $L$ is the one-dimensional normalization volume. In analogy to
the three-dimensional case $\alpha$ is a dimensionless coupling
constant. In the following we choose units such that
$\hbar=m_0=\omega=1$. With these assumptions the perturbation theory
produces the ground state energy $E_g = -\alpha$ and the inverse
effective mass $1/m = 1 - \alpha/2$.

{\it RG method}.
In the RG we consider the $S$-matrix
\begin{eqnarray}
S &=& T e^{-i\int dt\,H(t)} \nonumber\\ 
&=& e^{-i\int dt\, H_0} T e^{-i\int dt\,H_1(t)_I}\,,
\label{Seq}
\end{eqnarray}
where $H_1(t)_I$ is the interaction part of the Hamiltonian taken in
the interaction picture with respect to $H_0$. The idea of the RG is
to leave the $S$-matrix invariant while successively integrating out
diagrams of different time scales. The procedure is schematically
shown in Fig.~\ref{rgpl}. For a given cutoff $t_c$ in time
space, we allow only for correlation functions of the phonons with a
time scale $t>t_c$. At zero temperature, the latter are given by
\begin{eqnarray}
\langle (a^{\dagger}_{-q} + a_q)(t) (a^{\dagger}_q + a_{-q})\rangle
= e^{-i\omega t}\,.
\label{gameq}
\end{eqnarray} 
All correlation functions with time scales shorter than $t_c$ are accounted 
for by renormalized energies and coupling constants.
A change of $t_c$ to $t_c+dt_c$ is made by applying three steps
\cite{schoeller1}: 
(i) expanding the second exponential in (\ref{Seq}) and introducing normal
ordering for the phonon operators using Wick's theorem, (ii) integrating 
over the contractions with a time scale between $t_c$ and $t_c+dt_c$
and (iii) resumming the operators in an exponential
form. Consequently these operators will not be limited to zero-phonon and
one-phonon operators any more. But we shall see that a good
approximation is achieved for not too large coupling constants if we neglect
double or higher order vertex operators. One advantage of this method is
that we only need to consider the operators of the electronic system,
as the phonon degrees of freedom are integrated out in each
$t_c$-step. For the polaron problem we start with $t_c=0$ involving 
the operators as in
(\ref{H0eq}), (\ref{H1eq}) and end up with an effective Hamiltonian for
$t_c\rightarrow\infty$. We obtain the following RG-equations
\begin{eqnarray}
 \frac{d\epsilon_k}{dt_c} &=& -i \sum_q e^{-i\Delta_k^q t_c}
M_{k+q}^{-q}M_k^q
\label{depseq} \\
 \frac{dM_k^q}{dt_c} &=& \sum_{q'} \Biggl(M_{k+q+q'}^{-q'}M_{k+q'}^q
M_k^{q'} \frac{e^{-i\Delta_{k+q}^{q'}t_c} -
  e^{-i\Delta_k^{q'}t_c}}{i(\Delta_{k+q}^{q'} - \Delta_k^{q'})}
\Biggr. \nonumber\\ 
 && \Biggl. \hspace{1cm}+\,\,M_k^q M_{k+q'}^{-q'} M_k^{q'} t_c
e^{-i\Delta_k^{q'}t_c}\Biggr)\,.
\label{dMeq}
\end{eqnarray}
Here we introduced
\[ \Delta_k^q = \epsilon_{k+q} - \epsilon_k + \omega\,.\]
The $(k,q)$-dependence of the interaction coefficients is generated
during the RG-flow. The second term in (\ref{dMeq}) is a
correction term. It is due to the fact that a time interval
connected with a contraction becomes a single point in time at one
RG step. In the next step this leads to the generation of new terms
which were previously not present.
The level broadening is also included in (\ref{depseq}) since all
energies become complex.
The terms that generate the double vertex operators are of fourth
order in $M$. Therefore the equations (\ref{depseq}) and (\ref{dMeq}) contain
the order $M^4$ exactly.

{\it Feynman's method}.
To compare the results of the real-time RG we applied Feyman's method
to the one-dimensional polaron with a finite band-width. Feynman used a
variational principle in the path-integral formalism. Within his
two-particle approximation one obtains for the ground state energy
\begin{eqnarray}
 E_g &=& \frac{(v-w)^2}{4v} -
\alpha\frac{v}{\sqrt{\pi}}\int_0^{\infty}d\tau\,
e^{-\tau} \nonumber\\ 
 && \times [g(\tau)]^{-\frac{1}{2}}
[{\rm erf}(\frac{Dg(\tau)}{v^2})^{\frac{1}{2}}]
\label{Egfeyeq}
\end{eqnarray}
with the band-width $D$, the error-function
\[ {\rm erf(x)} = \frac{2}{\sqrt{\pi}}\int_0^xdt\,e^{-t^2}\]
and
\[ g(\tau) = \left(\frac{v^2-w^2}{v}\left(1-e^{-v\tau}\right) +
  w^2\tau\right)\,.\]
$v$ and $w$ are chosen such that $E_g$ is minimum.
For $v=w$ the result of the perturbation theory of first order in
$\alpha$ is reproduced. Following Feynman we treat small couplings by
setting $v=(1+\delta)w$ and $w=1$ \cite{com1}. 
Considering $\delta$ small one can now expand the
right-hand-side in (\ref{Egfeyeq}) and minimize it with respect to
$\delta$. The resulting energy is
\begin{eqnarray}
 E_g &=& -\alpha\frac{2}{\pi}\arctan{\sqrt{D}} 
 - \alpha^2\frac{4}{\pi^2}\left(3\arctan{\sqrt{D}}\right. 
\nonumber\\
 && \left. - 2\sqrt{2}\arctan{\sqrt{\frac{D}{2}}} -
  \frac{\sqrt{D}}{D+1}\right)^2
\label{feyn-en}\,.
\end{eqnarray}
Within Feynman's approximation we obtain for
the effective mass:
\begin{eqnarray}
 m &=& 1 + \frac{\alpha}{\pi}\left(\arctan{\sqrt{D}} +
  \sqrt{D}\frac{D-1}{(D+1)^2}\right) 
\nonumber\\
 && + \alpha^2\frac{4}{\pi^2}\left(3\arctan{\sqrt{D}}
 - 2\sqrt{2}\arctan{\sqrt{\frac{D}{2}}} -
  \frac{\sqrt{D}}{D+1}\right) \nonumber \\
 && \times \left(-3\arctan{\sqrt{D}} +
  3\sqrt{2}\arctan{\sqrt{\frac{D}{2}}} \right. \nonumber\\ 
  && \left. \hspace{1cm}+\,\,
\sqrt{D}^3\frac{3D^3+18D^2+21
D-2}{(D+1)^3(D+2)^2}\right)
\label{feyn-mass}\,.
\end{eqnarray}
Since (\ref{Egfeyeq}) follows from a mininum principle Feynman's
method leads to more accurate results for the energy than for the mass
(for quantitative studies concerning the accuracy in three dimensions
see \cite{lu}).

Our results for the energy using Feynman's approach as well as our
values following from perturbation theory of second order are in good
agreement with \cite{devreese}.

{\it Results}.
The differential equations (\ref{depseq}) and (\ref{dMeq}) are solved
numerically. Regarding the oscillating terms in these equations one
recognizes that given a certain discretization in $q$-space one obtains
large errors with increasing frequency $t_c$. To avoid this the phase
$\Delta_k^q t_c$ has been interpolated in $q$-space.

Unfortunately (\ref{depseq}) and (\ref{dMeq}) do not show a convergent
behaviour for the ground state energy for $t_c\rightarrow\infty$. One
reason is that there are undamped modes corresponding to high
excitations in the $q$-sums leading to
increasing effects on the ground state energy. In this context another
problem arises from the fact that the correlation function in
(\ref{gameq}) is not decaying. As a consequence,
oscillations decay as a function of $t_c$ but reoccur
for sufficiently large $t_c$ \cite{com2}.
The idea of our solution is to neglect further renormalizaton
effects of both $\Delta_k^q$ and $M_k^q$ for $t_c$ larger than
a certain point $t_f$. By doing this we obtained a damped oscillation
of the ground state energy $\epsilon_0$ in $t_c$-space for $t_c>t_f$. Therefore
(\ref{depseq}) can be integrated analytically which leads to
\begin{eqnarray}
 E_g &\equiv& \lim_{t_c\rightarrow\infty}\epsilon_0(t_c) \nonumber\\
&=& \epsilon_0(t_f)  - \sum_q
\frac{M_{q}^{-q}(t_f)M_0^q(t_f)}{\Delta_0^q(t_f)}e^{-i\Delta_0^q(t_f)t_f}\,.
\label{etfeq}
\end{eqnarray}

\begin{figure}
\centerline{\psfig{figure=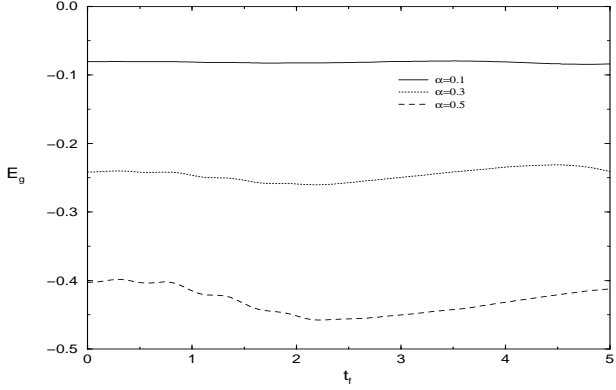,width=8cm,height=5.2cm,angle=-90}}
\caption{Ground state energy of the polaron in the renormalization 
group method as a function of $t_f$ for $D=10$. Solid line:
$\alpha=0.1$. Dotted line: $\alpha=0.3$. Dashed line: $\alpha=0.5$.}
\label{etfpl}
\end{figure}

In Fig.~\ref{etfpl} the solution of (\ref{etfeq}) is shown for
different $t_f$. The problems mentioned above make it
necessary to choose a finite $t_f$ where the renormalization effects
beyond perturbation theory are contained but the numerical instabilities
do not yet occur. We choose $t_f=2.5$ for all values of $\alpha$.
At this point only low 
excitations ($\Delta<0.4$) are not integrated out yet. Since 
$\Delta\sim 1$ sets the scale for the first excited state, it is reasonable
to assume that excited states do not have further important 
renormalization effects. The change of $E_g$ between $t_f=2$ and 
$t_f=3$ is approximately $2\%$ for $\alpha=0.5$.

\begin{figure}
\centerline{\psfig{figure=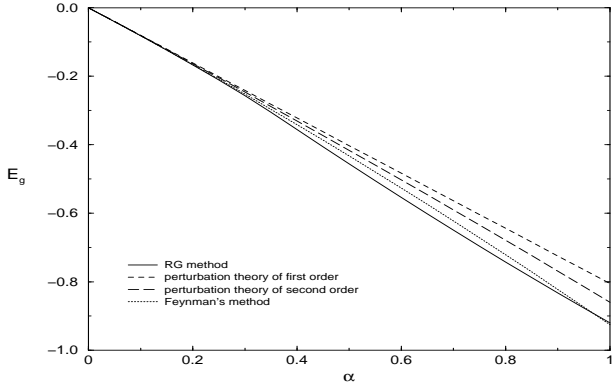,width=8cm,height=5.2cm,angle=-90}}
\caption{Ground state energy of the polaron as a function
of $\alpha$ for $D=10$. Solid line: Renormalization group method
described in this paper with $t_f=2.5$. Dashed and long-dashed line:
Perturbation theory of order $\alpha$ and $\alpha^2$. Dotted line: 
Feynman's method.}
\label{ealphapl}
\end{figure}

The $\alpha$-dependence of $E_g$ is shown in Fig.~\ref{ealphapl}. 
One notices that for
$\alpha<0.7$ we obtain lower values for the
ground state energy than those of Feynman's method. For larger
couplings our method is no longer reliable and yields worse results
(see Fig.~\ref{ealphapl}) which is not surprising since
we neglected double and higher order vertex operators.

\begin{figure}
\centerline{\psfig{figure=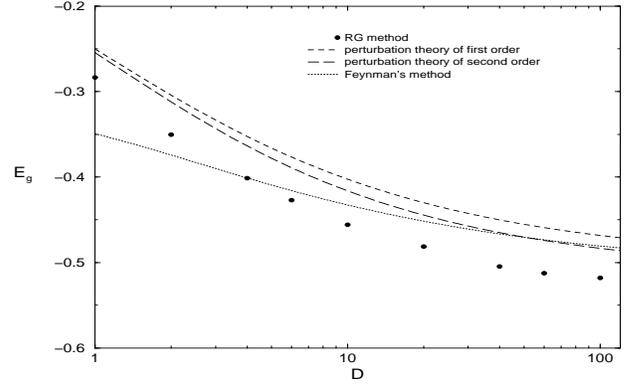,width=8cm,height=5.2cm,angle=-90}}
\caption{Ground state energy of the polaron as a function
of $D$ for $\alpha=0.5$. Circles: Renormalization group method
described in this paper with $t_f=2.5$. Dashed and long-dashed line:
Perturbation theory of order $\alpha$ and $\alpha^2$. Dotted line: 
Feynman's method.}
\label{ebpl}
\end{figure}

We also calculated $E_g$ for different band-widths $D$. As one can see
in Fig.~\ref{ebpl} for larger band-widths the deviation
to the results of Feynman's method grows.

Another quantity of interest is the mass of the polaron. If one
differentiates (\ref{depseq}) twice with respect to $k$ one
obtains a differential equation for $1/m$. However, its solution 
oscillates with increasing amplitude. Therefore for determining the
mass we calculate both the ground state energy and low excited
energies using the same procedure as above. The resulting dispersion
gives us a value for the mass. 
The accuracy for the mass is worse
though. For $\alpha=0.3$ the change of $1/m$ is approximately $6\%$
between $t_f=2$ and $t_f=3$.

\begin{figure}
\centerline{\psfig{figure=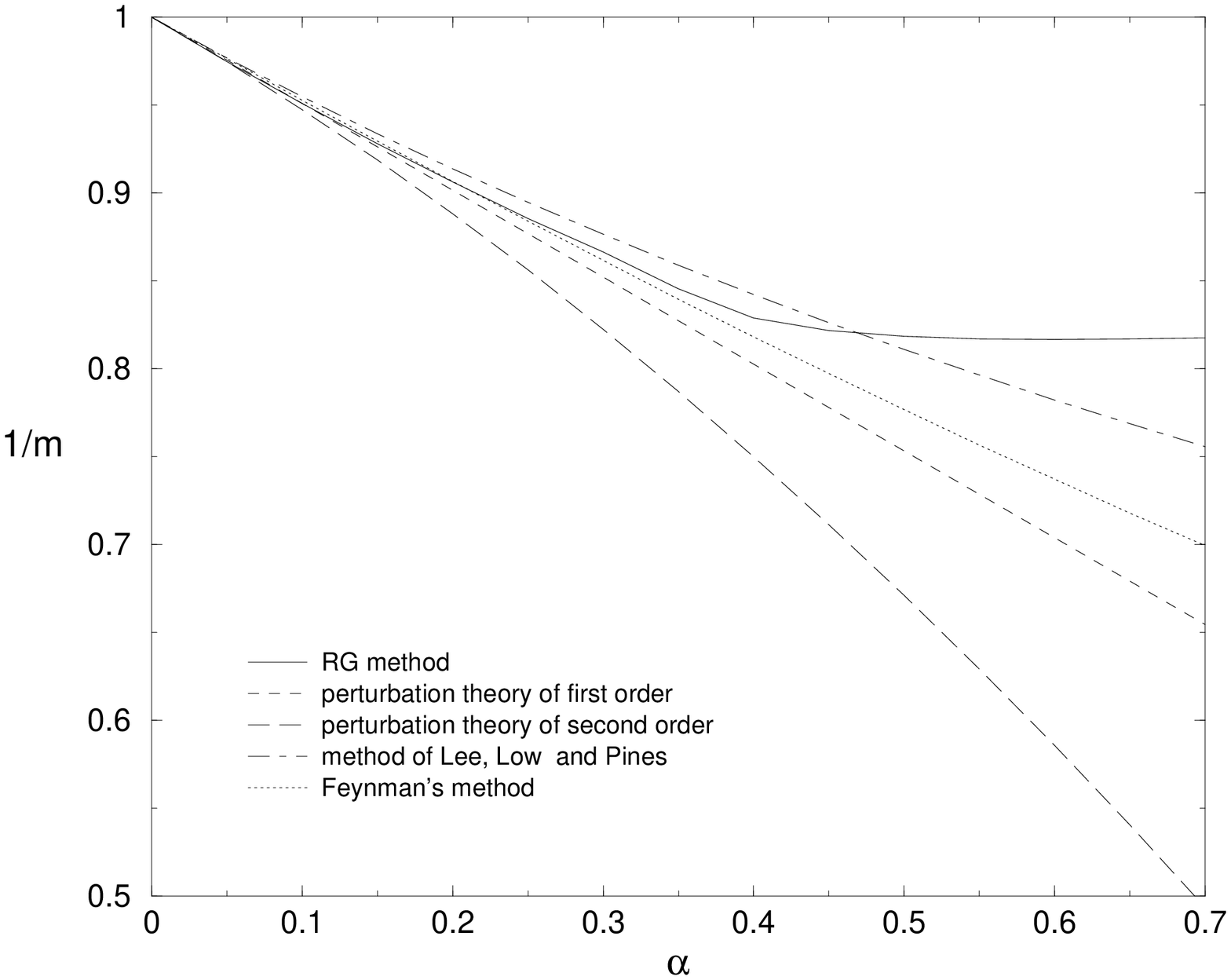,width=8cm,height=5.2cm,angle=-90}}
\caption{Inverse mass of the polaron as a function
of $\alpha$ for $D=10$. Solid line: Renormalization group method
described in this paper with $t_f=2.5$. Dashed and long-dashed line:
Perturbation theory of order $\alpha$ and $\alpha^2$. Dotted line: 
Feynman's method. Dot-dashed line: result of Lee, Low and Pines.}
\label{malphapl}
\end{figure}

\begin{figure}
\centerline{\psfig{figure=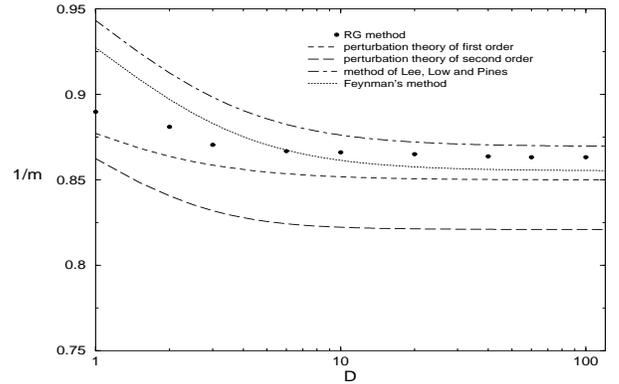,width=8cm,height=5.2cm,angle=-90}}
\caption{Inverse mass of the polaron as a function
of $D$ for $\alpha=0.3$. Circles: Renormalization group method
described in this paper with $t_f=2.5$. Dashed and long-dashed line:
Perturbation theory of order $\alpha$ and $\alpha^2$. Dotted line: 
Feynman's method. Dot-dashed line: result of Lee, Low and Pines.}
\label{mbpl}
\end{figure}

The results for different $\alpha$ are shown in Fig.~\ref{malphapl}. 
For small couplings ($\alpha<0.4$), we find a value for the mass 
between the variational principle of Feynman and the one of
Lee, Low and Pines \cite{com3}. For larger couplings the numerical 
solution is too unstable to make definite statements from the
RG-approach. From Fig.~\ref{mbpl} we see that our mass depends 
only slightly on the band-width. 


In summary, we used a new renormalization group method to study the
polaron problem. Although there arise some difficulties from the
structure of the flow-equations, we were able to investigate the
energy beyond perturbation theory. The mass was calculated as well,
but its accuracy is worse than that of the energy. The results
were compared to the values following from perturbation theory,
Feynman's method and the one of Lee, Low and Pines.
The performance of the RG is good for couplings $\alpha\lesssim 0.5$. 
This restriction is due to the neglecting of double and higher
order vertex operators and the undamped oscillating correlation
function of the phonons.

We acknowledge useful discussions with A. Mielke and K. Sch\"onhammer.
This work was supported by the ''Deutsche Forschungsgemeinschaft'' 
as part of ''SFB 345'' (M.K.) and ``SFB 195'' (H.S.)

\begin{figure}
\centerline{\psfig{figure=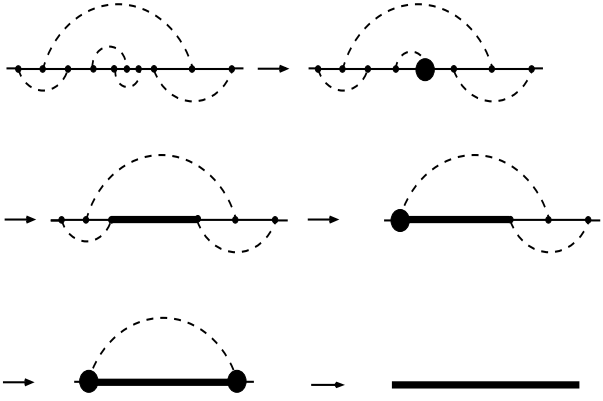,width=8cm,height=5.2cm,angle=0}}
\caption{Scheme of the renormalization group.}
\label{rgpl}
\end{figure}

\begin{references}
\bibitem{com1}
Feynman also minimizes the
energy with respect to $w$, but states that $E_g$ is not sensitive
to the choice of $w$. We found that in case of finite band-width
there is no minimum  with respect to $w$ within the validity of the
expansion.
\bibitem{com2}
We expect the numerics to be much more stable for other 
correlation functions, e.g. for acoustic phonons.
\end{references}
\end{document}